# Effect of pseudo datasets for the classification-based engineering design

Xianping Du [1*], Kai Zhang[1], Onur Bilgen[1], Laurent Burlion[1], and Hongyi Xu[2*]

[1] *Department of Mechanical and Aerospace Engineering, Rutgers University, Piscataway, NJ 08854, USA*

[2] *Department of Mechanical Engineering, University of Connecticut, Storrs, CT 06269, USA*

## Abstract:

Machine learning classification techniques have been used widely to recognize the feasible design domain and discover hidden patterns in engineering design. An accurate classification model needs a large dataset; however, generating a large dataset is costly for complex simulation-based problems. After training by a small dataset, surrogate models can generate a large pseudo dataset efficiently. Errors, however, may be introduced by surrogate modeling. This paper investigates the mutual effect of a large pseudo dataset and surrogate modeling uncertainty. Four widely used methods, i.e., Naive Bayes classifier, support vector machine, random forest regression, and artificial neural network for classification, are studied on four benchmark problems. Kriging is used as the basic surrogate model method. The results show that a large pseudo dataset improves the classification accuracy, which depends on both design problems and classification algorithms. Except for the Naive Bayes, the other three methods are recommended for using pseudo data to improve classification performance. Also, a wind turbine design problem is used to illustrate the effect of the pseudo dataset on feasible subspace recognition. The large pseudo dataset improves the recognized subspace bound greatly, which can be reproduced by classification models well except for the Naive Bayes classifier. Under the uncertainty of surrogate modeling, the random forest presents high robustness to recognize the feasible design domain bound, while the artificial neural network demonstrates a high sensitivity to uncertainty with the recognized bound deteriorated.

## 1. Introduction

Machine learning (ML) has been widely adopted to extract knowledge from large datasets in, e.g., finance, natural language processing, and health sciences. In engineering, increasing attention has also been given to its application in data-driven design [1]. Most of the design practices can be categorized as forwarding or inverse mapping

[*] Corresponding author:
1. Xianping Du, Department of Mechanical and Aerospace Engineering, Rutgers University, 98 Brett Rd, Piscataway, NJ 08854, USA
2. Hongyi Xu, Department of Mechanical Engineering, University of Connecticut, Storrs, CT 06269, USA
E-*mail*: Xianping Du (xianping.du@rutgers.edu); Hongyi Xu (hongyi.3.xu@uconn.edu)

[2]. For a design problem with design variable vector $x$ and responses $y$, the forward mapping takes the values of all design variables to predict their response after learning the internal rules ($y = f(x)$) between them from data [3]. The regression techniques are suitable for this type of problem. Nevertheless, inverse mapping seeks to recognize a comprehensive set of designs or design subspaces where output prerequisites can be satisfied [4].

The ML classification method can classify the different combinations of design variables into several performance classes after learning their interrelation from data. Consequently, it can be utilized to recognize a full set of designs or subspaces based on response performance requirements for inverse mapping [5-7], reliability-based design [8-11], and complex system interpretation [12-15]. As a mathematical model, a trained classifier gains the ability to explore the design space with low computational cost. Thus, a set of designs or a subspace with expected performance can be recognized efficiently. By generating a set of high-performance designs, the clustering and decision tree-based methods were used for the design of airfoils [7], bridges [6], and vehicle energy-absorbing component [5]. Also, due to the capability to recognize the subspaces efficiently, the reliability-based design or analysis can be enhanced in the reliability analysis of structural or high-dimensional systems [9, 10], the recognition of important design regions from implicit constraints in sequential sampling [8], and reliability-based design with time-variant probabilistic constraints [11]. Furthermore, classification methods can also help interpret the complex system by learning from data. Aiming at high safety performance, the vehicle structural system was modeled by a classification technique, that is, the decision tree. A multi-level design framework was developed to model the whole design problem at multiple levels, that is, system-component-subcomponent levels. The design rules were further interpreted at multiple levels [5, 12, 14]. Furthermore, uncertainty induced by the variations of design variables was also modeled and qualified quite efficiently by a new decision tree technique for learning from uncertain data [13, 15].

To implement the classification method in engineering designs, a highly accurate and robust classification model is expected. This can often be achieved or improved by using a large and high-fidelity dataset generated by simulations, which could provide enough information regarding a specific problem. A large and high-fidelity dataset could describe the objective problem well and thus, refine the recognized subspace boundaries by inverse mapping. However, generating a large dataset by simulations is expensive for many engineering problems, for example, the vehicle crashworthiness finite element simulation [13], and airplane computational fluid dynamic (CFD) simulations [16]. A method to reduce the cost of data generation without much quality compromise is required.

Generating the pseudo data is an efficient way to enhance the model performance, which is also called data augmentation in machine learning. A variety of methods can be used for data augmentation, for example, adding noise, resampling, and segmentation [23]. These methods generate pseudo data by operating the real data. Model accuracy and performance can be improved by an enlarged training dataset after adding pseudo data. In engineering design, the surrogate model method is often used to generate pseudo data to enhance the computational cost of expensive simulations in, e.g., multidisciplinary optimization, and uncertainty quantification [17, 18]. There are several widely used regression algorithms for surrogate modeling, e.g., the response surface model (polynomial function), Gaussian process regression (GPR, also known as Kriging), artificial neural network (ANN), and radial basis function [19]. After training by data, the relationship between design variables and responses can be approximated by a surrogate model, which calculates the response of new samples efficiently. It could generate a large pseudo dataset to enhance the classification tasks. A large dataset benefits the classification for exploring the design space. However, errors may be induced by the surrogate modeling at unseen design regions [20], which may negate the benefits of a large dataset.

It is non-trivial to investigate the effect of a large pseudo dataset and the uncertainty due to surrogate modeling. The effect of the large pseudo dataset can be represented by either the classification accuracy variations or further the precision changes of the recognized subspace [21], whereas it is difficult to quantify the uncertainty induced by the surrogate model and its propagation. A promising method is needed to represent the uncertainty induced by surrogate modeling and its propagation to the generated design sets or recognized subspaces in classification-based inverse mapping. Regarding the surrogate model uncertainty, many studies are available in different areas of engineering design including multidisciplinary optimization, hierarchical system design, and reliability-based design [22]. Thus, the uncertainty also needs to be considered in classification-based inverse mapping practice. The mutual effect of the large pseudo dataset and surrogate modeling uncertainty has not been studied.

**Contributions**

In this paper, an investigation is conducted to deepen our understanding of the effect of pseudo datasets on the classification-based engineering design. Four widely used machine learning (ML) methods, Naive Bayes classifier (NBC), support vector machine (SVM), random forest (RF), and artificial neural network (ANN) are selected as the representative classification techniques. Four benchmark problems are selected as examples. The Kriging model is used as the basic surrogate model algorithm to generate a large pseudo dataset for each problem. The following investigations are made:

1) The effect of the large pseudo dataset is studied by comparing the classification accuracy of the small physical dataset and large pseudo datasets.
2) The dataset size effect of physical and pseudo datasets are analyzed for applying the knowledge of this study to problems with different available sizes of physical datasets.
3) Furthermore, an engineering problem for a wind turbine design is designated for discussing the effect of large pseudo datasets on the recognized bounds of feasible subspace and the variations due to uncertainty.

**Outline**

The manuscript is organized as follows: Section 2 introduces the theoretical background. In Section 3, the design methodology is presented. In Section 4, the details are introduced including the design problems, surrogate modeling, pseudo datasets generation, hyperparameters optimization, and classification modeling. The results are analyzed in Section 5 for the four benchmark problems. In Section 6, the effect of the pseudo data on the recognized subspace bounds is discussed. Conclusions are drawn in Section 7.

# 2. Theoretical background

## 2.1 Pseudo data generation by surrogate model and uncertainty

In this paper, the Kriging method, also known as Gaussian process regression, is used as the basic surrogate model for generating pseudo data due to its wide application, high accuracy, and low training cost [25]. Its basic architecture can be expressed by polynomial and stochastic terms as follows,

$$\boldsymbol{y} = f(\boldsymbol{x}) + \delta(\boldsymbol{x})\,, \tag{1}$$

where, $f(\boldsymbol{x})$ is a polynomial regression model and $\delta(\boldsymbol{x})$ is a zero-mean stochastic term. The $f(\boldsymbol{x})$ represents the global trend of the problems and is composed by $N_p$ polynomial terms ($\boldsymbol{p}(\boldsymbol{x})$) with a weight vector $\alpha_i$ for them and expressed as,

$$f(\boldsymbol{x}) = \boldsymbol{\alpha} \cdot \boldsymbol{P}(\boldsymbol{x}) = \sum_{i=1}^{N_p} \alpha_i \cdot p_i(\boldsymbol{x})\,, \tag{2}$$

$\delta(\boldsymbol{x})$ is a measure of local variation and achieved by a stochastic process with zero mean and $\sigma^2$ process variance and nonzero covariance. Its covariance matrix can be presented as,

$$\boldsymbol{cov}[\delta(x_i), \delta(x_j)] = \sigma^2 \boldsymbol{R}(k(x_i, x_j))\,, \tag{3}$$

where $\boldsymbol{R}$ is the covariance matrix, which is symmetric and positive definite with one on diagonal. and $k(x_i, x_j)$ is the correlation function to approximate the correlation between any two data points $x_i$, and $x_j$. The Gaussian correlation is frequently used and expressed as,

$$k(x, x') = e^{-\boldsymbol{\eta}\langle x_i, x_j\rangle^2}\,. \tag{4}$$

where, $\langle x_i, x_j\rangle$ is the Euclidean distance between two samples, $x_i$ and $x_j$. The $\boldsymbol{\eta}$ is the correlation parameter vector that needs to be fitted by data. Also, the polynomial coefficient $\boldsymbol{\alpha}$ and variance $\sigma^2$ can also be estimated from training data by maximum likelihood estimation. After that, the model can predict the mean ($\mu^*$) and response variance ($\sigma_y^{*2}$) at any new locations ($\boldsymbol{x}^*$) by,

$$\mu^* = f(\boldsymbol{x}^*) + \boldsymbol{r^T R^{-1}}(\boldsymbol{y} - \boldsymbol{f}(\boldsymbol{x})), \tag{5}$$

$$\sigma_y^{*2} = \boldsymbol{\sigma^2}(\boldsymbol{1} - \boldsymbol{r^T R^{-1} r} + (\boldsymbol{1} - \boldsymbol{P^T R^{-1} r})/\boldsymbol{P^T R^{-1} P}). \tag{6}$$

The $\boldsymbol{r}$ is the correlation vector between the new locations and the training data points. The root mean square error (RMSE) and mean absolute error ($MAE$) are used as the accuracy measures, which are expressed by,

$$RMSE = \sqrt{\frac{\sum_{i=1}^{n}(y_i - y_i^*)}{n}}\,, \tag{7}$$

$$MAE = \frac{\sum_{i=1}^{n}|y_i - y_i^*|}{n}, \tag{8}$$

where $y$ and $y^*$ are the real and predicted responses, respectively; $n$ is the number of samples.

Based on the Kriging prediction, at any new data point, the estimated values follow the normal distribution $y^* \sim N(\mu^*, \sigma_y^{*2})$. In this way, the uncertainty from the estimation at the new locations can be approximated by this normal distribution. Meanwhile, since the approximation is based on the Euclidean distance of new points to training points, more data and close to training point may cause a lower uncertainty and so versa. Thus, increasing the number of training data points could reduce the degree of uncertainty in a random sampling of test data while the computational cost may also be increased dramatically. The surrogate modeling uncertainty at unknown areas has to be accounted for, especially under a small number of training data.

## 2.2 Classification with hyperparameters optimization

Classification is a type of supervised learning technique, which predicts the label of new data after training. In this way, after learning from engineering designs with design variables ($\boldsymbol{x}$) and response labels ($L$), e.g., good ( $L_g$) or poor labels ($L_p$), where the 'g' and 'p' represent the good and poor designs. A design alternative can be labeled as $L_g$ or $L_p$ based on a response threshold by,

$$\text{Determine } L = L_g \text{ if } f(\boldsymbol{x}) \leq y_c\text{; otherwise } L = L_p, \tag{9}$$

where $y_c$ represent the critical response values for the good or poor design and $f(\boldsymbol{x})$ is the response function. The classification method can learn the interrelation between the design variables and response classes from labeled data. Then, for the new design ($\boldsymbol{x}^*$), its label distribution (probabilities of 'g' and 'p' labels for good and poor designs, respectively) can be predicted by,

$$\text{Determine } L = L_g \text{ if } p\left(L_g \middle| \boldsymbol{x}^*\right) > p\left(L_p \middle| \boldsymbol{x}^*\right)\text{; otherwise } L = L_p. \tag{10}$$

In this study, four well-known machine learning algorithms, that is, NBC, SVM, RF, and ANN, are selected as the basic algorithms for investigating the pseudo data effect on their performance in classification-based engineering designs. The algorithm details are well introduced in the literature [2, 13, 25, 26] and we do not repeat them here.

The architecture and performance of a machine learning model depend on both hyperparameters and model parameters. Model parameters are trained by data but hyperparameters need to be determined by users before training [13, 19, 23]. In this study, a sequential model-based method is used for hyperparameter optimization by the framework developed in [19].

The loss functions need to be defined for assessing the classification performance. Before introducing loss functions, the confusion matrix is explained as shown in Figure 1, which is a table regarding 'g' and 'p' labels. For a labeled dataset, this table summarizes the number of 'g' samples that are predicted as 'g' ($N(Tg)$) and 'p' ($N(Fp)$) and the number of 'p' that is predicted as 'g' ($N(Fg)$) and 'p' ($N(Tp)$).

| Classes | | Predicted | |
|---|---|---|---|
| | | 'g' | 'p' |
| True | 'g' | $N(Tg)$ | $N(Fp)$ |
| | 'p' | $N(Fg)$ | $N(Tp)$ |

Figure 1 Confusion matrix for the two classes: 'g' and 'p'

The most important measure is the classification accuracy ($ACC$) as expressed in Eq. (11), which assesses the global accuracy.

$$ACC = \frac{N(Tg)+N(Tp)}{N}, \tag{11}$$

where $N$ is the total number of samples. In addition to $ACC$, the $precision$ is adapted as calculated by,

$$precision = \frac{N(Tg)}{N(Tg)+N(Fg)}, \tag{12}$$

which suggests the ability of the model to generate a set of 'g' designs so a high $precision$ is preferred. Meanwhile, the ability of a model to recognize all 'g' areas in the design space is also considered. The $TPR$, also called $recall$ is taken as expressed by,

$$TPR = recall = \frac{N(Tg)}{N(Tg)+N(Fp)}. \tag{13}$$

A bigger $TPR$ is expected to recognize all 'g' subspaces from the whole design space. Finally, the $F1$ score evaluates the balance between the $precision$ and $recall$ n,

$$F1 = 2 * \frac{precision*recall}{precision+recall}. \tag{14}$$

$ACC$ is used as the optimization objective and all the above measures are used to evaluate model performance.

### 2.3 Relationship between the surrogate and classification models

As the main purpose of this study is to investigate the impact of using a pseudo dataset on the performance of classification models, here we discuss two ways of classifying the design space into sub-regions based on the design performances.

(1) **Regression-based method**. With a surrogate model trained on a small physical simulation dataset, we can recognize a design subspace by setting a threshold on the surrogate model predictions. The accuracy of the surrogate model determines the accuracy of the classification boundaries.

(2) **Classification-based method.** Another way is to train a classification model on a large pseudo dataset generated by the surrogate model. In the following sections, we are going to investigate whether using pseudo data can improve the accuracy of the classification model. On one hand, surrogate modeling may induce uncertainty to pseudo data. On another hand, the classification model may act as a "regularizer" that resolves the possible over-fitting issue in the surrogate model.

## 3. Methodology for pseudo data effect study

The methodology of this study is diagrammed in Figure 2 with four steps included, i.e., the physical dataset generation, pseudo datasets generation by the surrogate model, classification modeling, and the pseudo data effect analysis.

**Step 1** starts the framework by defining the design problems, that is, the response, design variables, and constraints, etc. A physical dataset is generated by the design of the experiments (DOE) using the Latin hypercube method (LHM) to distribute the sampling points in the design space as evenly as possible. Their responses are calculated by simulations. This physical dataset is partitioned into the physical training and test dataset ($\Psi$) with a specific ratio. Six physical training subsets (**Ω**) are resampled from the training dataset for the data size effect study.

**Step 2** aims to generate pseudo data. Kriging models are trained by the **Ω**, which is tested by the $\Psi$. DOE is also implemented by LHM with the response predicted by the Kriging model, which forms six large pseudo datasets corresponding to each subset in **Ω**. Eight subsets are resampled from each large pseudo dataset to form the pseudo subsets (**Λ**).

The classification performance of pseudo data is tested in **Step 3**. This contains two principal parts. On one hand, the classification performance of the regression-based method is tested by comparing the labeled Kriging-predicted response and real response on Ψ. On the other hand, **Λ** is used to train classification models, which are tested by Ψ in the classification-based method. Before implementing these two methods, hyperparameter optimization is adopted with 5 folds cross-validation. The optimization reduces the effect of the uncertainty induced by manually or experience-based tunning. Using the optimum, the classifiers can be constructed and trained by either **Ω** or **Λ**.

**Step 4** analyzes the pseudo dataset effect on the classification accuracy, the design subspace recognition, and uncertainty propagation. The dataset size effect is also studied.

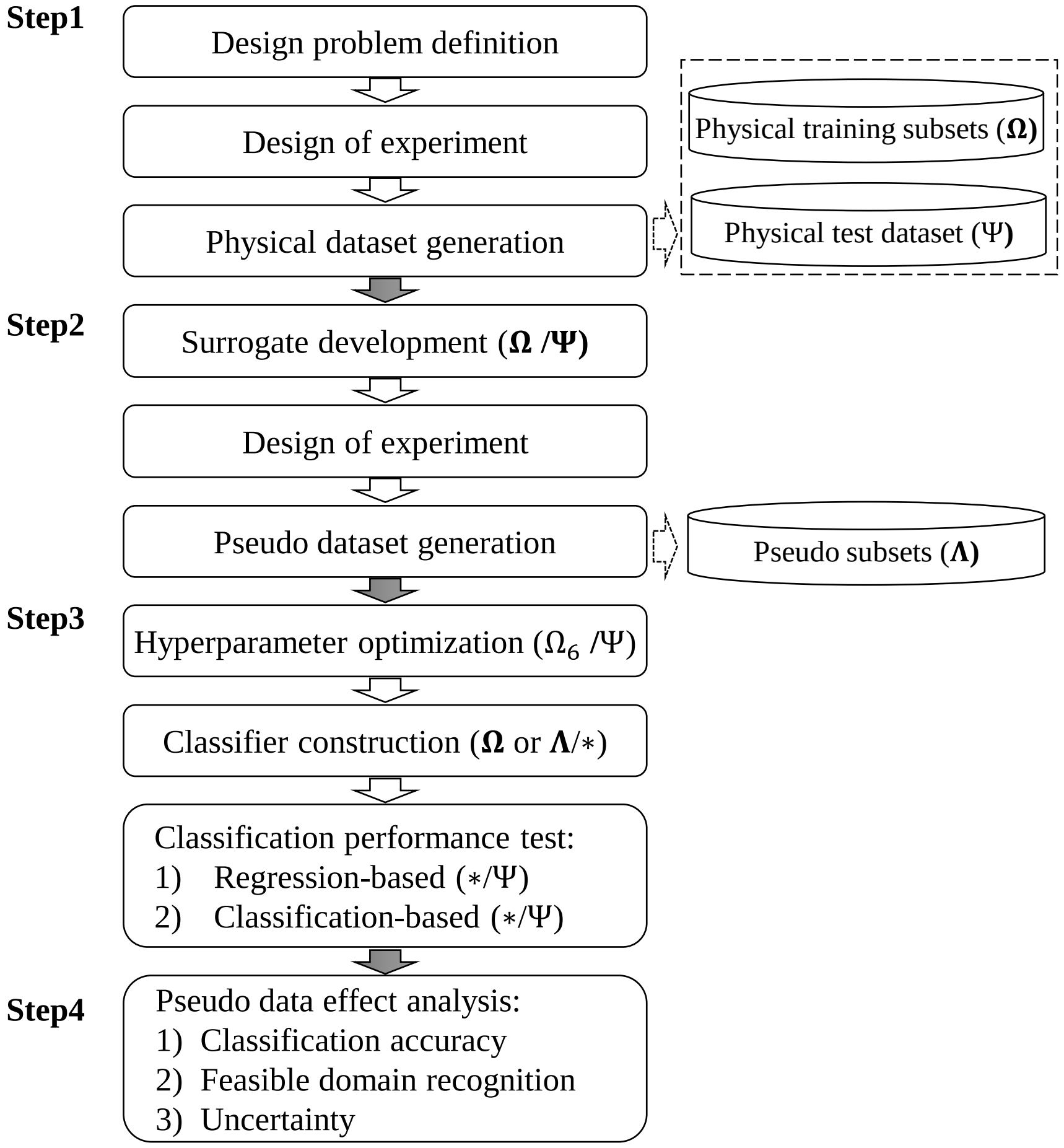

Figure 2 Workflow with four steps to study the effect of the large pseudo dataset with different datasets denoted with characters; the datasets used by a process is also noted by these characteristics in brackets; For example, '$\Omega_6$ /$\Psi$' means the process with the 6th physical training subset for training and physical test dataset for test and '∗/Ψ' suggests only the test process with Ψ and so versa for '**Ω** or **Λ**/∗'.

# 4. Introduction of case studies, data generation, and ML model training

Four widely used benchmark problems are used as examples. The effect of large pseudo datasets is analyzed on classification accuracy. Furthermore, a computationally expensive problem, i.e., wind turbine design, is introduced and used only for studying the effect of the pseudo dataset and surrogate modeling uncertainty to the feasible space recognition.

## 4.1 Design problem introduction

### *4.1.1 Benchmark problems from the literature*

The four benchmark problems from literature are summarized in Table 1. The *Wing Weight* problem is to calculate the weight ($m_w$) of an airplane wing by a mathematical function with ten independent parameters [29]. The *U-beam* problem is a U-shaped structure with seven independently structural and load parameters and a mathematical function is developed to calculate the maximum effective strain ($\varepsilon_{max}$) of the structure [15]. The *Pressure Vessel* problem predicts the cost ($\mathcal{H}$) of a pressure vessel with material, forming, and welding costs considered [32]. The *Welding Beam* problem describes the fabrication cost ($\mathcal{L}$) of welding a beam to a substrate with heavy constraints applied [33]. The details of these problems are provided in the literature and are not repeated here.

Table 1 Characteristics of the four mathematical problems

| No. | Problem | Response (unit) | No. of variables | Variable type | Design domain | Reference |
|---|---|---|---|---|---|---|
| **1** | *Wing Weight* | $m_w$ | 10 | Mixed* | Continuous | [29] |
| **2** | *U-beam* | $\varepsilon_{max}$ | 7 | Mixed | Continuous | [15] |
| **3** | *Pressure Vessel* | $\mathcal{H}$ | 4 | Single | Discontinuous | [32] |
| **4** | *Weld Beam* | $\mathcal{L}$ () | 4 | Single | Discontinuous | [33] |

Note: *Mixed: variables with different units, e.g., 'm' and 'kg'

For each physical problem, the Latin Hypercube Method (LHM) is used to sample 1,500 points in the feasible design space, which could explore the design space as much as possible using limited data points. Their responses are calculated by the corresponding function and form the physical datasets. Each physical dataset is partitioned into training and test datasets with a ratio of 2:1 (1,000 and 500 points, respectively). The response distributions of the physical training and test sets are shown in Figure 3. The test dataset has almost the same distribution as its corresponding training dataset, which suggests both the training and test datasets can describe their corresponding design problem well.

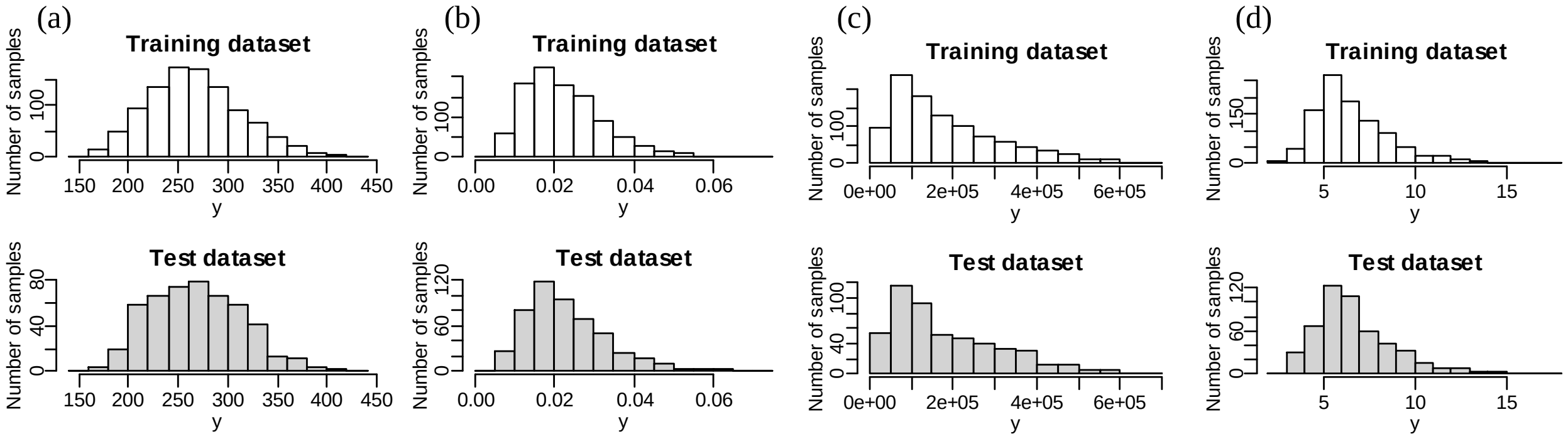

Figure 3 Distribution of the physical training and test datasets for problem (a) *Wing Weight*, (b) *U-beam*, (c) *Pressure Vessel*, and (d) *Weld Beam*.

*4.1.2 DTU 10MW wind turbine power coefficient (DTU10MWPC) problem*

A wind turbine extracts energy from wind and is becoming a common renewable energy conversion method. The rotor is the most critical component for power generation as shown in Figure 4. In this paper, the DTU 10MW is a horizontal axis turbine with 10MW rated power, which was developed by the Technical University of Denmark [34] and is used as an example. The rotor diameter and hub height are 178.3 m and 119 m, respectively. The details of the turbine can be seen in [34, 35]. For a turbine, the power coefficient determines the efficiency of power extraction as expressed by,

$$P = P_a \cdot C_p, \tag{15}$$

where, $P$, $P_a$, and $C_p$ are the rotor extracted power, wind power, and power coefficient, respectively. $P_a$ can be represented as: $P_a = 0.5\rho A v^3$, where $\rho$, $A$, and $v$ are the air density, rotor swept area, and wind speed, respectively. $C_p$ is depends on the blade tip speed ratio ($\lambda$) and blade pitch angle ($\beta$) as expressed by,

$$C_p = C_p(\lambda, \beta). \tag{16}$$

The $\lambda$ is the ratio of blade tip speed to wind speed and can be expressed as, $\lambda = R\omega/v$, where $R$, and $\omega$ are the rotor radius and rotational speed, respectively. $\beta$ is the blade pitch angle, which is adjusted by rotating around its longitudinal axes. Their mutual impact to $C_p$ is shown in Figure 4 with $\lambda$ changing between 3 and 14.75 and $\beta$ varying from 0° to 24.75°. This mutual effect is important for modern wind turbine control to track the power generation [36].

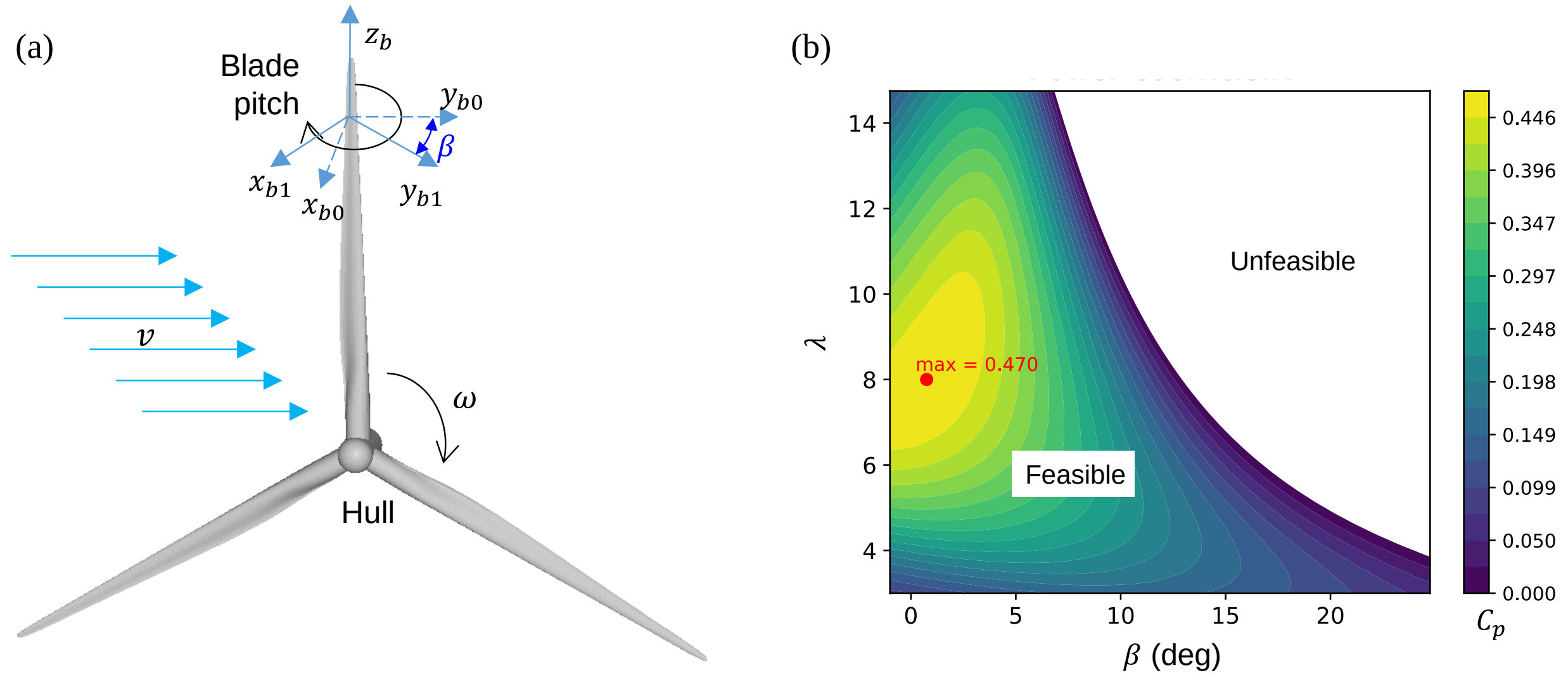

Figure 4 (a) DTU 10MW reference wind turbine rotor with its rotation and blade pitch angle ($\beta$) denoted and (b) the simple diagram shows the dependency of $C_p$ on $\lambda$ and $\beta$ with the maximum $C_p$

The $C_p$ is often computed by expensive simulations, e.g., by CFD solvers. In the $\lambda$ and $\beta$ space, there is an unfeasible domain. To recognize the feasible domain can help to evaluate the model only at the feasible area to reduce the computational cost. In this paper, we try to recognize the feasible $C_p$ (>0) bound. The DOE is conducted in the design space with 40 combinations of $\lambda$ and $\beta$ generated by the LHM. They are partitioned into physical training and test datasets equally as shown in Figure 5(a).

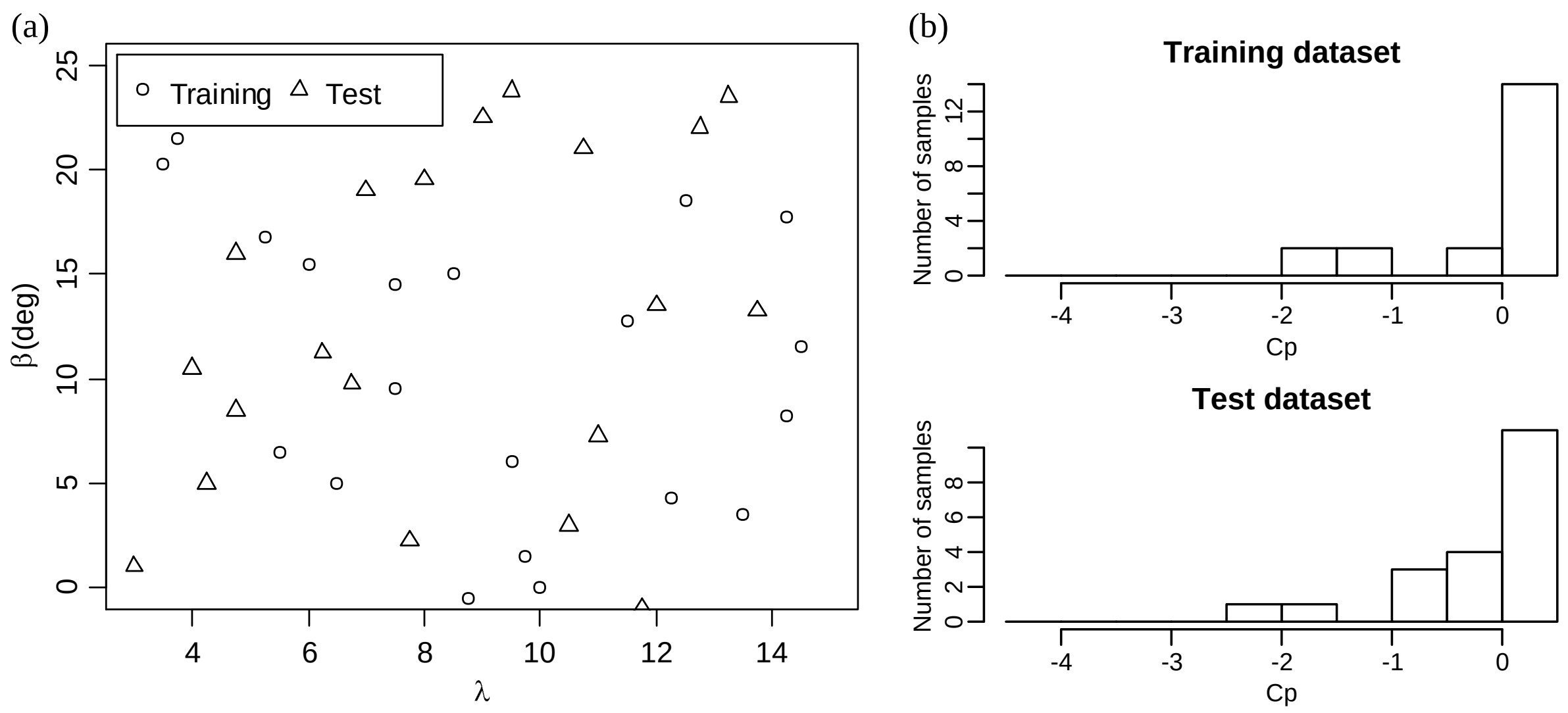

Figure 5 (a) $C_p$ values of the 40 points and (b) $C_p$ distribution in physical training and test datasets

The $C_p$ values are evaluated by an open-source wind turbine simulator called OpenFAST [37]. The $C_p$ distribution are shown in Figure 5(b). There is similar but poor distribution due to the small dataset size. It should be noted that this problem is only used to analyze the pseudo data effect on the design subspace identification in Section 6.

**4.2 Pseudo datasets generation by surrogate models**

For each of the four benchmark problems, Kriging models are trained by the physical training subsets. To study dataset size effect, six subsets with different sizes (20, 50, 100, 200, 500, and 1,000) are resampled with replacement from the whole physical training dataset. The trained Kriging models are tested by the corresponding *physical test dataset* with the RMSE and MAE shown in Appendix A. With the increase of subset sizes, the errors are reduced greatly. Also, the predicted value is paired with the corresponding real value on the *physical test dataset*. They are plotted with a slope one perfect match (PM) line in Figure 6. The closer the point to the PM line, the better prediction is achieved. As expected, increasing the dataset size improves the degree of matching. The *Wing Weight* problem converges slower than the others as increasing training dataset sizes. Finally, increasing the dataset size raises the surrogate model accuracy.

For each design problem, 50,000 points are sampled by LHM in the design space with the response evaluated by the corresponding Kriging model. This forms the six pseudo datasets. Each pseudo dataset is resampled into eight pseudo subsets 1 to 8 with replacement (100, 200, 500, 1,000, 2,000, 5,000, 10,000, and 50,000 points). Thus, 48 (6 surrogate model * 8 pseudo subsets) pseudo subsets are generated for each problem.

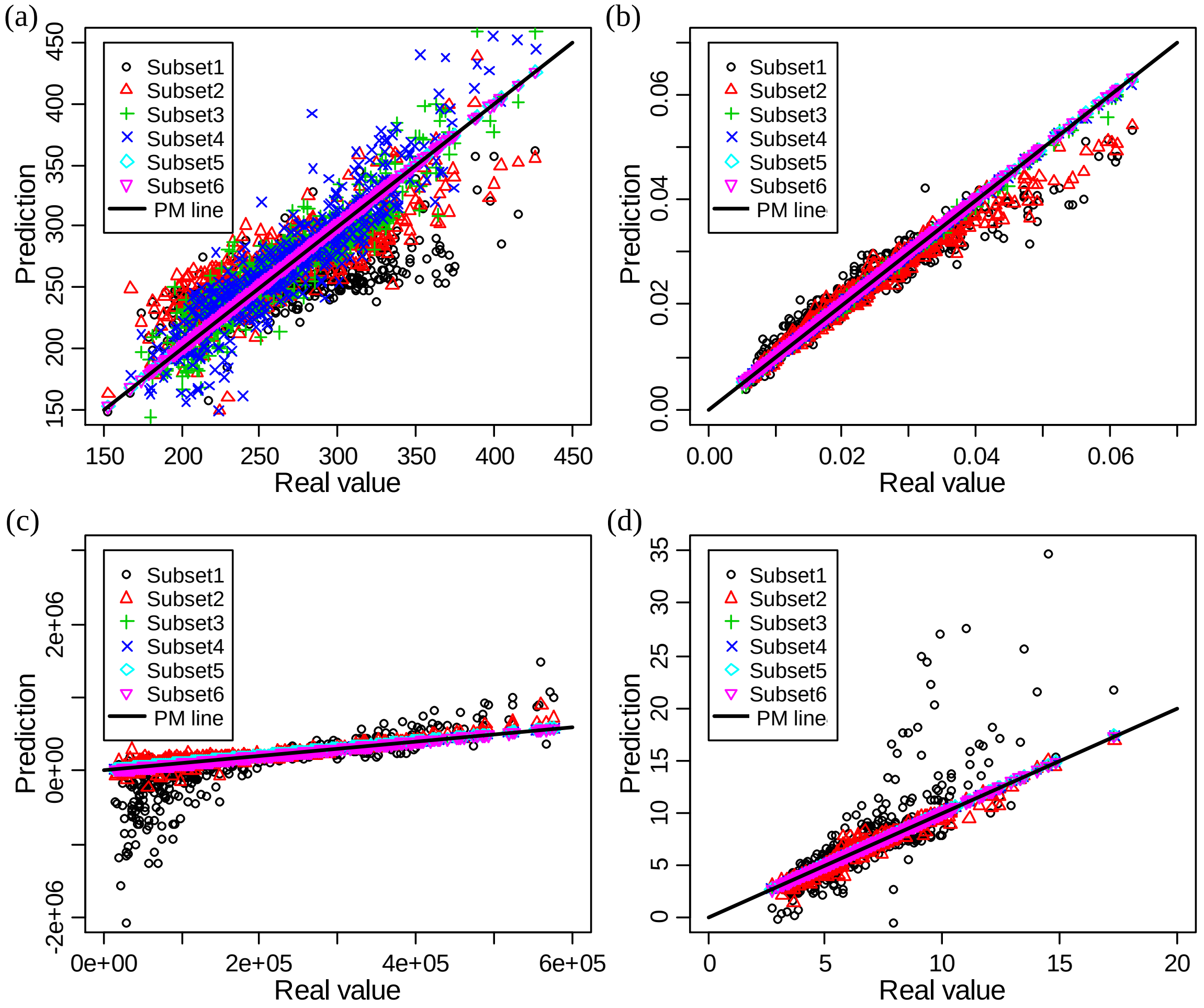

Figure 6 Surrogate model prediction against real values on the *physical test dataset* with a PM line for the six physical training subsets (Subset1 to 6) of (a) *Wing Weight*, (b) *U-beam*, (c) *Pressure Vessel*, and (d) *Weld Beam*

### 4.3 Hyperparameter optimization

The whole physical training and test datasets are then labeled with the approximate ratio of good ('g') and poor ('p') to be 1:1. The response thresholds and resulted label ratios are listed in Table 2.

Table 2 Label ratio of 'g' to 'p' in physical training and test dataset with thresholds

| **Problem** | **'g' response threshold** | **Label 'g' to 'p' ratio in physical datasets** | |
|---|---|---|---|
| | | **Training (1,000 points)** | **Test (500 points)** |
| ***Wing Weight*** | $\leq$ 264.9 | 1.020 | 0.961 |
| ***U-beam*** | $\leq$ 0.02095 | 1.020 | 0.961 |
| ***Pressure Vessel*** | $\leq$ 142, 044.7 | 1.000 | 1.000 |
| ***Weld Beam*** | $\leq$ 6.134 | 1.045 | 0.916 |

The hyperparameters are optimized [25] with 5 folds cross-validation. The ML models are then constructed with the optimal values of hyperparameters. These models together with the NBC are trained by the whole physical training dataset and tested by the *physical test dataset* with a high accuracy shown in Table 3.

Table 3 Test $ACC$ of classifiers after hyperparameter optimization

| **MLA** | **Problem** | | | |
|---|---|---|---|---|
| | ***Wing Weight*** | ***U-beam*** | ***Pressure Vessel*** | ***Weld Beam*** |
| **NBC** | 0.920 | 0.942 | 0.924 | 0.862 |
| **SVM** | 0.978 | 0.992 | 0.996 | 0.980 |
| **RF** | 0.912 | 0.944 | 0.960 | 0.918 |
| **ANN** | 0.980 | 0.984 | 0.994 | 0.960 |

The pseudo data effect is then investigated. To test the classification accuracy of the regression-based method, its prediction is labeled using thresholds in Table 2, and compared with the response labels of the *physical test dataset*. Furthermore, in the classification-based method, classification models are trained by either the physical training or corresponding pseudo subsets with the optimal hyperparameter values. They are tested by the *physical test dataset*. $ACC$, $precision$, $TPR$, $F1$ are calculated to demonstrate different abilities to explore the design space.

### 4.4 Pseudo data effect analysis

For each benchmark problem, the pseudo data effect is analyzed regarding three aspects: (1) on regression-based method; (2) on classification-based method; (3) dataset size effect. The baseline classification model is trained by the smallest physical training subset (20 points). For each problem, the baseline model accuracy is compared with the Kriging-based classification accuracy for implementing (1). Also, (2) is completed by comparing the baseline model with the whole pseudo dataset trained classifiers on their accuracy. Including the different sizes of physical training and pseudo subsets in (1) and (2), (3) can be investigated.

In addition, the pseudo data effect is analyzed on the feasible space recognition using the DTU10MWPC problem. The feasible subspace bound recognized by the classifiers trained by a small physical training set is compared with the ones identified by regression-based and classification-based methods. Also, the effect of uncertainty is propagated by pseudo data to study its effect on subspace recognition.

## 5. Results of the classification accuracy of benchmark problems

This section analyzes the three aspects listed in Section 4.4 using the four benchmark problems. To demonstrate the effect of a large pseudo dataset, the baseline classification models are compared with models by the whole pseudo dataset in Sections 5.1 and 5.2. The other subsets are only used to discuss the dataset size effect in Section 5.3.

### 5.1 Pseudo dataset effect on classification accuracy by labeling surrogate model response

To demonstrate the pseudo dataset effect by regression-based method, the classification accuracy of Kriging trained models is calculated on the *physical test dataset* and compared with the accuracy of the baseline classification models in Figure 7.

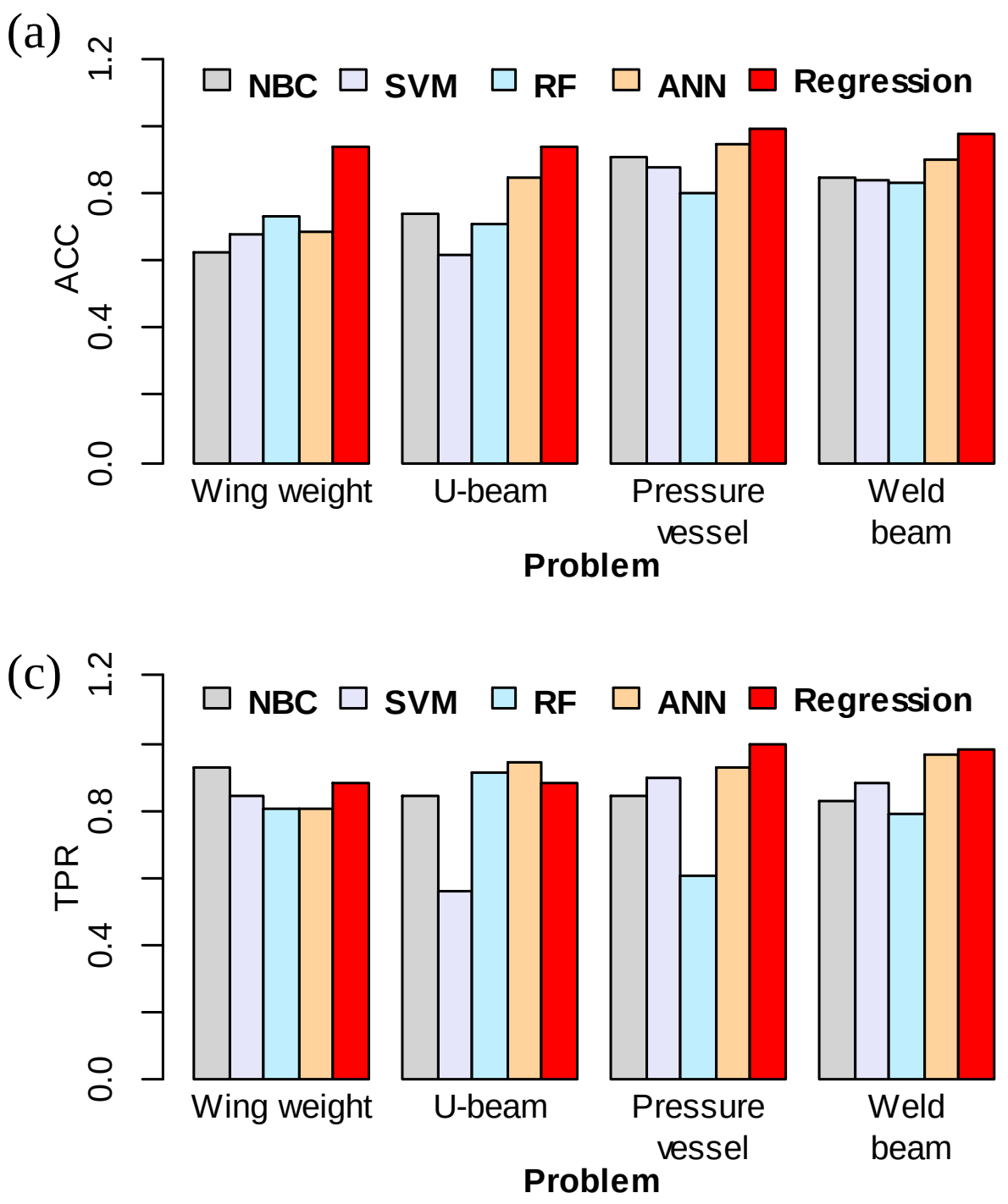

Figure 7 Pseudo data effect on classification accuracy of (a) $ACC$, (b) $precision$, (c) $TPR$, and (d) $F1$, by comparing the regression-based and classification-based methods

Most of the cases present the accuracy improvement by the regression-based method. Only a small portion (3/64) of cases presented lower performance of the large pseudo dataset. This suggests that the surrogate model can produce more information for improving classification tasks than the original sparse physical dataset.

## 5.2 Large pseudo dataset effect on classification

Based on the generated large pseudo dataset, their effect on the classification-based method is also investigated. The accuracy of the baseline classification models is compared with the these trained by corresponding whole pseudo dataset (50,000 points). The results are summarized in Table 4 with the change ($\delta$) of $ACC$ calculated in percentage, where a positive value suggests an increase. The mean and S.D. of $ACC$ and $\delta$ are calculated among four problems and four ML models at the bottom and right of each portion, respectively.

Generally, a large pseudo dataset gains a better $ACC$ than its corresponding small physical training dataset, which is suggested by the positive increment of $ACC$ made by all 16 cases. In detail, the maximum $ACC$ increment reaches 51.3% and the mean $\delta$ is larger than 10% for four ML methods. Also, the improvement depends on the different design problems and ML methods, and the former presents a higher impact than the latter. This can be demonstrated by the relatively larger S.D. of $ACC$ $\delta$ among different problems than its among different ML models. It seems the discontinuous design spaces of Problem 3 and 4 impede the $ACC$ improvement due to the relatively lower $\delta$. For different ML methods, the SVM presents the highest $ACC$ $\delta$ but also a high S.D. (17.1%) while the RF presents the most robust performance by the lowest S.D. In general, the SVM, RF, and ANN models improve $ACC$ more than NBC.

Table 4 ACC comparison of classifiers trained by Physical training subset 1 and corresponding Pseudo subset 8 (Note: Negative $\delta$ are denoted with the **bold**)

| Measure | Problem | NBC | | | SVM | | | RF | | | ANN | | | Mean | S.D. |
|---|---|---|---|---|---|---|---|---|---|---|---|---|---|---|---|
| | | physical | pseudo | $\delta$/% | physical | pseudo | $\delta$/% | physical | pseudo | $\delta$/% | physical | pseudo | $\delta$/% | $\delta$/ % * | $\delta$/% * |
| $ACC$ | 1 | 0.624 | 0.894 | 43.3 | 0.682 | 0.936 | 37.2 | 0.736 | 0.936 | 27.2 | 0.690 | 0.936 | 35.7 | 35.9 | 5.7 |
| | 2 | 0.742 | 0.914 | 23.2 | 0.620 | 0.938 | 51.3 | 0.710 | 0.928 | 30.7 | 0.848 | 0.940 | 10.8 | 29.0 | 14.7 |
| | 3 | 0.906 | 0.926 | 2.2 | 0.880 | 0.990 | 12.5 | 0.800 | 0.978 | 22.3 | 0.950 | 0.990 | 4.2 | 10.3 | 7.9 |
| | 4 | 0.844 | 0.862 | 2.1 | 0.838 | 0.972 | 16.0 | 0.830 | 0.964 | 16.1 | 0.902 | 0.986 | 9.3 | 10.9 | 5.8 |
| | Mean $\delta$/ % + | --# | -- | 17.7 | -- | -- | 29.3 | -- | -- | 24.1 | -- | -- | 15.0 | -- | -- |
| | S.D. $\delta$/ + | -- | -- | 17.1 | -- | -- | 15.9 | -- | -- | 5.5 | -- | -- | 12.2 | -- | -- |
| $precision$ | 1 | 0.571 | 0.925 | 61.8 | 0.632 | 0.986 | 56.1 | 0.700 | 0.978 | 39.7 | 0.648 | 0.982 | 51.5 | 52.3 | 8.1 |
| | 2 | 0.695 | 0.990 | 42.6 | 0.624 | 0.991 | 58.7 | 0.645 | 0.991 | 53.7 | 0.788 | 0.991 | 25.7 | 45.2 | 12.7 |
| | 3 | 0.968 | 0.914 | **-5.5** | 0.865 | 0.984 | 13.7 | 0.987 | 0.996 | 0.9 | 0.971 | 0.988 | 1.8 | 2.7 | 6.9 |
| | 4 | 0.843 | 0.835 | **-0.9** | 0.802 | 0.959 | 19.7 | 0.847 | 0.962 | 13.6 | 0.852 | 0.979 | 15.0 | 11.9 | 7.7 |
| | Mean $\delta$/ % + | -- | -- | 24.5 | -- | -- | 37.1 | -- | -- | 27.0 | -- | -- | 23.5 | -- | -- |
| | S.D. $\delta$/ + | -- | -- | 28.6 | -- | -- | 20.5 | -- | -- | 20.8 | -- | -- | 18.2 | -- | -- |
| $TPR$ | 1 | 0.931 | 0.853 | **-8.3** | 0.841 | 0.882 | 4.9 | 0.808 | 0.890 | 10.1 | 0.804 | 0.886 | 10.2 | 4.2 | 7.5 |
| | 2 | 0.845 | 0.833 | **-1.4** | 0.563 | 0.882 | 56.5 | 0.910 | 0.861 | **-5.4** | 0.943 | 0.886 | **-6.1** | 10.9 | 26.4 |
| | 3 | 0.840 | 0.940 | 11.9 | 0.900 | 0.996 | 10.7 | 0.608 | 0.960 | 57.9 | 0.928 | 0.992 | 6.9 | 21.9 | 20.9 |
| | 4 | 0.828 | 0.887 | 7.1 | 0.879 | 0.983 | 11.9 | 0.787 | 0.962 | 22.3 | 0.962 | 0.992 | 3.0 | 11.1 | 7.2 |
| | Mean $\delta$/ % + | -- | -- | 2.3 | -- | -- | 21.0 | -- | -- | 21.2 | -- | -- | 3.5 | -- | -- |
| | S.D. $\delta$/ + | -- | -- | 7.8 | -- | -- | 20.7 | -- | -- | 23.3 | -- | -- | 6.1 | -- | -- |
| $F1$ | 1 | 0.708 | 0.887 | 25.3 | 0.722 | 0.931 | 29.0 | 0.750 | 0.932 | 24.2 | 0.718 | 0.931 | 29.8 | 27.1 | 2.4 |
| | 2 | 0.762 | 0.905 | 18.7 | 0.592 | 0.933 | 57.5 | 0.755 | 0.921 | 22.1 | 0.859 | 0.935 | 8.9 | 26.8 | 18.4 |
| | 3 | 0.899 | 0.927 | 3.1 | 0.882 | 0.990 | 12.2 | 0.752 | 0.978 | 29.9 | 0.949 | 0.990 | 4.3 | 12.4 | 10.7 |
| | 4 | 0.835 | 0.860 | 2.9 | 0.838 | 0.971 | 15.8 | 0.816 | 0.962 | 18.0 | 0.904 | 0.985 | 9.0 | 11.4 | 5.9 |
| | Mean $\delta$/ % + | -- | -- | 12.5 | -- | -- | 28.6 | -- | -- | 23.6 | -- | -- | 13.0 | -- | -- |
| | S.D. $\delta$/ + | -- | -- | 9.8 | -- | -- | 17.8 | -- | -- | 4.3 | -- | -- | 9.9 | -- | -- |

Note: +: The mean and S.D. of $\delta$ among the four problems; *: The mean and S.D. of $\delta$ among the four ML models; #: the items are not calculated;

In addition to the $ACC$, the other three measures are also calculated. They present a similar trend with the $ACC$ but slightly different for $TPR$. Compared with other measures, more models present lower $TPR$ performance or even negative $\delta$. $TPR$ is the ability of the model to recognize all good subspaces or designs. Compared with the $precision$ or $ACC$, $TPR$ performance is less significant since it is hard and unnecessary to recognize all good subspaces. Generally, the total trend of these 16 cases still presents an improvement for $TPR$, but the improvement by NBC seems less significant due to many negative $\delta$. To select the ML methods for using a large pseudo dataset for classification performance improvement, recommendations can be given:

- SVM is recommended but concerns should be given to the hyperparameter tuning in case of the high computational cost mentioned in [19].
- RF and ANN also gain a good model accuracy and pseudo data improvement, where RF is preferred due to its high robustness, but their training cost is increasing with the increased dataset size dramatically [19].
- NBC can also improve model performance but less significant than the others. Nevertheless, it needs less computational cost.

### 5.3 Dataset size effect

For different problems, they may have a varied number of available physical data. The size effect of the physical training set is implemented to investigate the influence of different number training data in Section 5.3.1. Furthermore, although a large pseudo dataset could improve the classification performance, it also increased the training cost. Investigating the pseudo dataset size effect could help to understand the classification accuracy trend with respect to the pseudo training dataset sizes and help select a large enough pseudo dataset under acceptable computational cost. This is completed in Section 5.3.2.

#### *5.3.1 Physical data size effect*

The four classification models trained by the physical training subsets and then their corresponding pseudo subsets are tested with the $ACC$ shown in Figure 8. The comparisons of other measures are presented in Appendix B. In Figure 8, with the increase of physical training subset sizes, the $ACC$ of trained classification models are also lifted accordingly, which is especially obvious in the low size area. This suggests a suitable size physical training set is existing to get the trade-off between the model accuracy and the cost of physical data generation. For example, 200 points for the NBC of Problem 1. These trends can be observed on another three measures but several cases, e.g., the $TPR$ of RF and ANN of Problem 2, may not strictly follow it. In general, a physical training dataset with 200 points can train a good classification model for problems with a low to intermediate (<20) number of design variables. However, to evaluate 200 cases may be still difficult for expensive problems so pseudo data is required.

With the increasing sizes of the physical training dataset, the $ACC$ of its corresponding full pseudo dataset trained ML models is also increased. This suggests, increasing the physical training dataset size enriches the information to describe the problem and improves classification performance. This, however, may have occasional exceptions, e.g., the NBC of Problems 3 and 4. The other three measures present a similar trend as shown in Appendix BAppendix **B**: . Thus, we need to try to get more physical data points under acceptable computational cost, but a small dataset is also applicable to generate a large pseudo dataset for an accurate classifier.

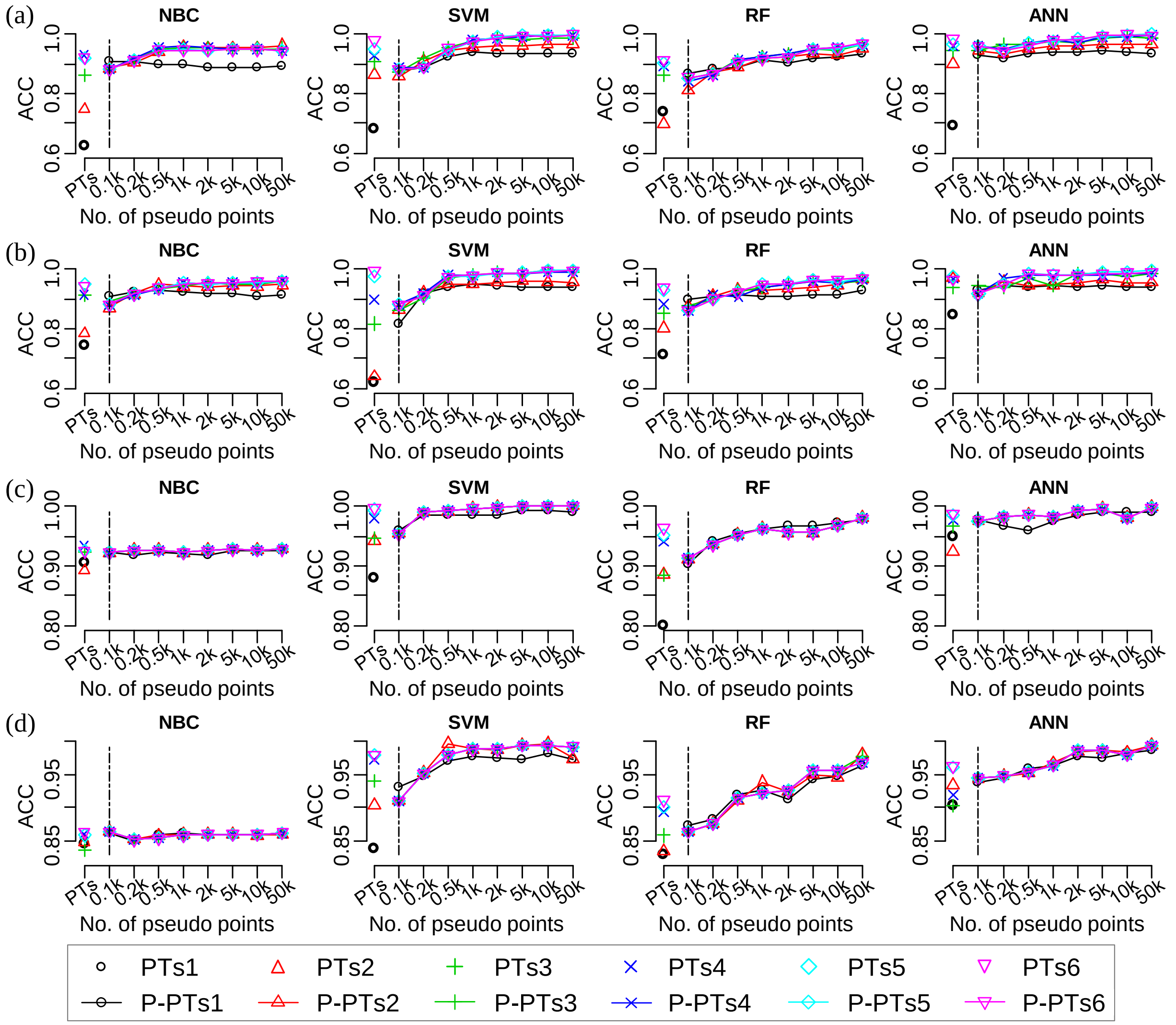

Figure 8 $ACC$ of four ML models trained by different size physical training subset and corresponding pseudo datasets for Problems (a) *Wing Weight*, (b) *U-beam*, (c) *Pressure Vessel*, and (d) *Weld Beam* (Note: P-PTs1: Pseudo subsets corresponding to Physical training subset 1 (PTs1))

### *5.3.2 Pseudo data size effect*

Increasing the pseudo dataset size could improve the $ACC$ of trained classifiers accordingly in a limited range. This suggests increasing the pseudo dataset could provide more information to model classification in addition to compensating the uncertainty caused by the surrogate modeling. This trend is more significant as the size of the pseudo dataset is less than, e.g., 1,000 (1k). This number seems also case-dependent. This knowledge is important because although pseudo data generation is cheap but to increase pseudo dataset size may increase the training cost of classifiers. Thus, increasing pseudo dataset size should consider the acceptable computation cost.

This trend is also sound even for the other three measures in Appendix B even several exceptions, e.g., the $ACC$ of the NBC of Problems 3 and 4, and the $precision$ and $TPR$ of the RF of Problem 4, which may present either an unchanged or slightly reduced accuracy measures by increasing pseudo dataset size.

# 6. Discussion of the classification boundary of the DTU10MWPC problem

In this section, the DTU10MWPC problem introduced in Section 4.1.2 is implemented for the feasible design subspace recognition. Based on the generated small physical training and test dataset, a Kriging model is trained and tested as presented in Appendix C. A large pseudo dataset with 50,000 points is generated by LHM and evaluated by the Kriging model. Four ML classification methods are trained by the physical training and pseudo training datasets, respectively. The test errors are compared in Appendix C. The regression-based and classification-based methods are used to recognize the bound of the feasible domain, where the $C_p$ threshold is zero.

## 6.1 Pseudo dataset effect on inverse mapping

The effect of the large pseudo dataset on the design subspace bound recognition is investigated. The bound recognized by the baseline classification model is compared with bounds from the regression-based and classification-based methods (with four ML models). All the recognized bounds, and the real bounds are shown in Figure 9 for comparison. The real contour of $C_p$ is also depicted for understanding the complex effect of $\lambda$ and $\beta$.

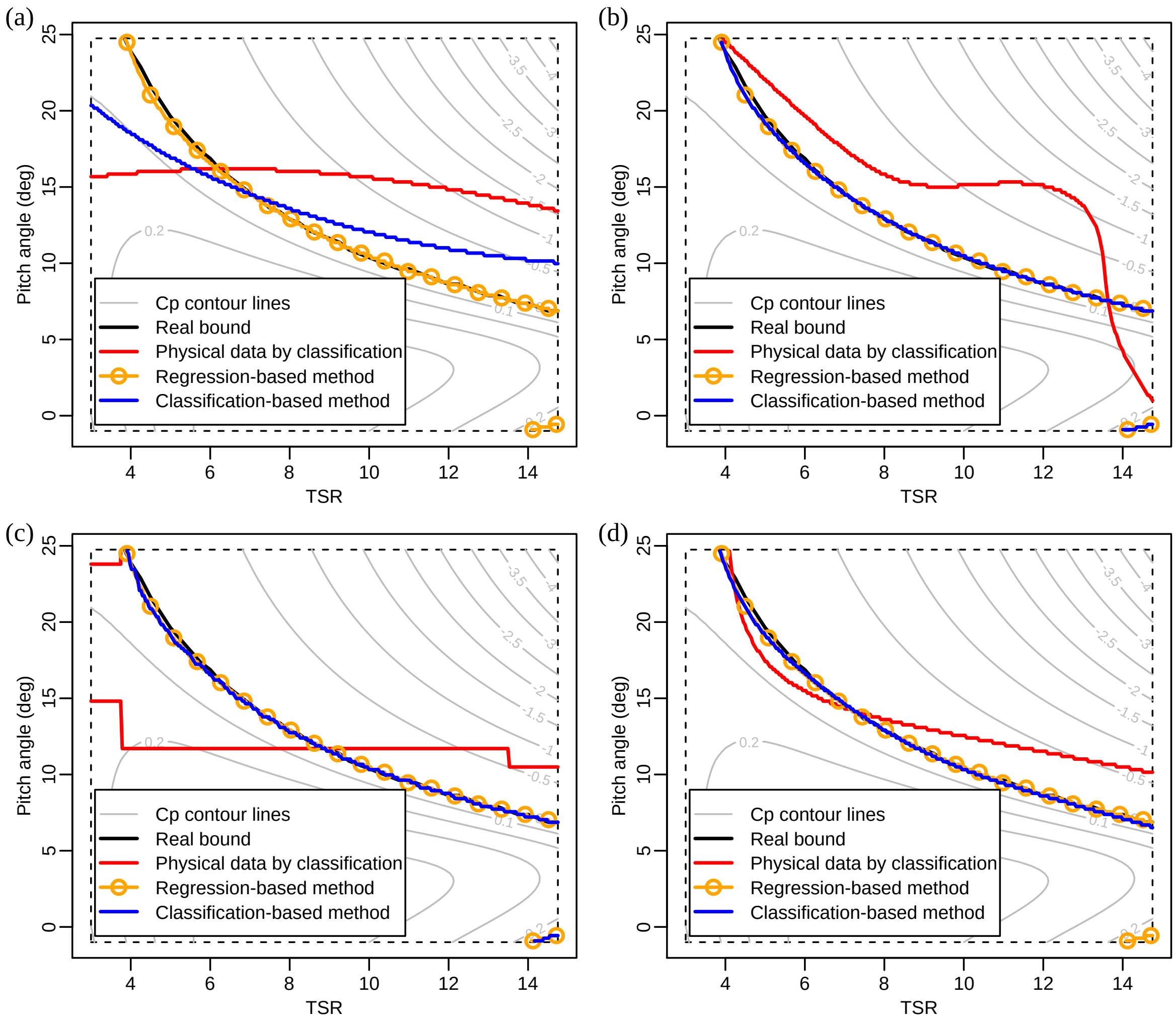

Figure 9 Effect of the large pseudo dataset for the feasible and unfeasible bounds by comparing the regression-based and classification-based method, that is, (a) NBC, (b) SVM, (c) RF, and (d) ANN

In Figure 9, all baseline models cannot recognize the bound and even the basic profile accurately, while the regression-based method recognizes the bound accurately but with a minor error at the lower-right corner. This suggests its limited ability to predict accurately at the boundary of design space due to the lack of enough information. This minor error, however, can be removed by the ANN model trained by pseudo dataset while SVM and RF reproduce the regression-based results. Even though the pseudo-data-trained NBC is not accurate enough, the profile shape and closeness to the real bound are improved from the physical training dataset. In this way, the ANN seems a good choice against this trend while the SVM and RF also perform well to recognize the bound.

**6.2 Uncertainty effect due to surrogate modeling**

The surrogate modeling uncertainty may be transferred to pseudo data, which then, impact the recognized subspace bounds. Using the Kriging, the uncertainty can be estimated at the new data points and represented by an interval with the upper and lower 95% bounds. This interval is used to represent the uncertainty of the surrogate model. The upper and lower 95% bound values of the 50,000 pseudo data points form two new pseudo datasets. The regression-based method is also implemented to recognize the bound. Classifiers are trained by the upper and lower 95% pseudo datasets accordingly after labeling for recognizing the 'g'/'p' bounds. The results of these two methods are compared with the real bound in Figure 10.

For the bound of the regression-based method, the lower 95% pseudo dataset takes greater impact with the main bound distorted. The error bound at the lower-right corner is moving toward different directions under upper and lower 95% pseudo datasets.

In Figure 10(a), the performance of NBC is not deteriorated due to the uncertainty. Although the classification model trained by the upper 95% pseudo dataset still recognizes an inaccurate bound, but the bound by the lower 95% pseudo dataset can match with the real bound well. This cannot conclude the uncertainty improves the NBC performance and may need more investigations.

A more general conclusion can be observed from Figure 10 (b) and (d), where the accurate bounds estimated by the classification-based method are distorted and moved due to the uncertainty, which follows the regression-based method. However, the ANN trained by the upper 95% recognizes the optimal area (near $\lambda$= 7.88 and $\beta$=0 deg) of $C_p$ as unfeasible, which suggests low robustness. RF in Figure 10(c) presents the best performance against the impact of uncertainty by keeping the bound shape unchanged. As a non-parametric method, the RF train the model by using the bagging technique to form a group of trees, which improves the robustness of the RF model [19, 27, 38].

The possible bound deteriorations are caused by the surrogate modeling trained by a small physical dataset, which induces a severe uncertainty on the unseen area. Regarding the performance against the side-effect of uncertainty, the RF presents the best performance with better robustness presented than the regression-based method and ANN is highly sensitive to the uncertainty, which may be caused by possible overfitting.

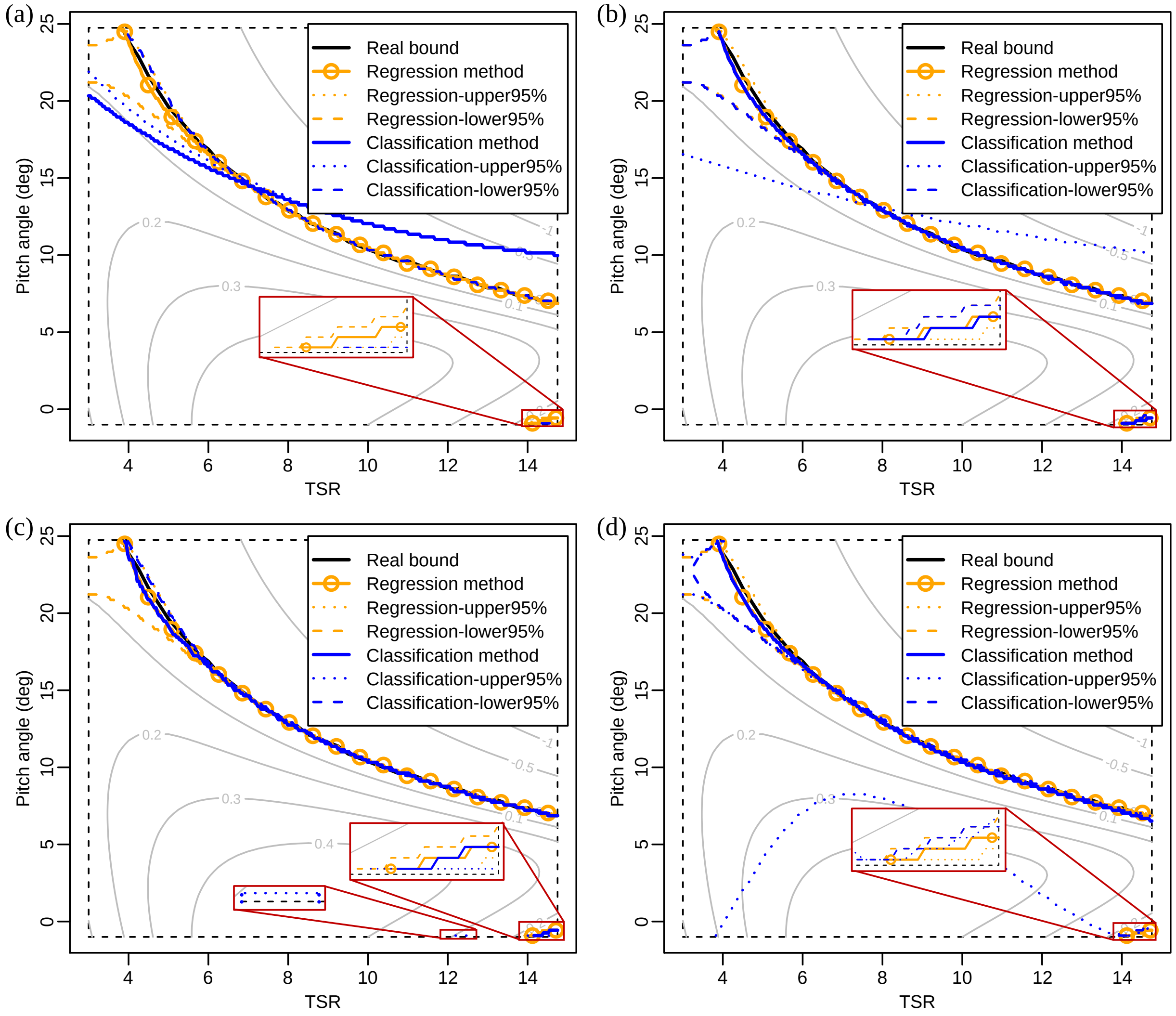

Figure 10 Bounds comparison between the regression-based method and the classification-based methods for (a) NBC, (b) SVM, (c) RF, and (d) ANN under different pseudo datasets to account for the uncertainty. (Note: the local views of some tiny areas at corners or boundaries are also shown.)

# 7. Conclusion

The classification method is often used in engineering inverse mapping. A large dataset is required to train an accurate model, which may be difficult for computationally expensive problems. A large pseudo dataset can be generated by a surrogate model after training by a relatively small physical dataset, but uncertainty may be introduced by surrogate modeling. In this paper, the effect of a large pseudo dataset on classification performance is investigated. Four well-known classification techniques are used in four benchmark problems. The classification accuracy of models trained by physical training subsets is compared with the corresponding pseudo datasets by the two methods: regression-based method by labeling pseudo data response and the classification-based method by training a classification model by labeled pseudo dataset. The pseudo dataset effect is discussed in three aspects: the regression-based method, classification-based method, and physical and pseudo dataset size effect.

The results show that the large pseudo dataset improves the classification accuracy greatly in the regression-based method. Accordingly, most classification models trained by a large pseudo dataset presented higher accuracy than the model trained by their small physical training subset, for example, the maximum $ACC$ increment of 51.2%. This demonstrates improvement made by the large pseudo dataset. The degree of improvement depends on the different design problems and classification algorithms where the former takes a higher impact. Also, SVM, RF, and ANN are recommended for their great improvement using pseudo datasets but the increased computational cost of RF and ANN under a large dataset needs to be considered.

In addition, the dataset size study shows that increasing the physical dataset size can improve the trained classifier accuracy, especially, at the relatively low size area. Nevertheless, a small physical dataset is also applicable to generate a high classification accuracy by producing a large pseudo dataset. This is helpful for computationally expensive problems. Increasing the pseudo dataset size gains the benefit of accuracy improvement within a limited degree. This, however, also cause the increased computational cost, which needs to be considered.

In addition, the pseudo dataset effect on the subspace bound recognition is studied. The results suggest a significant improvement by a large pseudo dataset on the accuracy of recognized bound. Especially, the SVM, RF, and ANN can re-produce the bound accurately. This could benefit the inverse mapping-based design and enhance the design space exploration.

Also, by incorporating the uncertainty into the pseudo dataset, the bound variations are uncovered. Under uncertainty, most of the recognized bounds are moved or distorted. These are mostly reproduced by the classification models. Under uncertainty, NBC-recognized bound does not deteriorate. Compared with the SVM and ANN, RF is highly robust to uncertainty due to its non-parametric characteristic. Although an investigation is completed in this study, more general conclusions still need deeper discussion in engineering design, e.g., improving the accuracy at the corners and problem dependency.

# Acknowledgments

This research is supported by the Department of Energy (DOE) Advanced Research Projects Agency-Energy (ARPA-E) Program award DE-AR0001186 entitled "Computationally Efficient Control Co-Design Optimization Framework with Mixed-Fidelity Fluid and Structure Analysis." The authors thank DOE ARPA-E Aerodynamic Turbines Lighter and Afloat with Nautical Technologies and Integrated Servo-control (ATLANTIS) Program led by Dr. Mario Garcia-Sanz. Special thanks to the entire ATLANTIS Team for their support. The authors are grateful for the computing resources at Amarel cluster provided through the Office of Advanced Research Computing (OARC) at Rutgers University.

# Conflict of interest

The authors declare that they have no known competing financial interests or personal relationships that could have appeared to influence the work reported in this paper.

## Replication of results

The results and codes of this study can be available per request for replication.

## Appendix A: Accuracy measures of surrogate models

Table A. 1 shows the test errors, i.e., RMSE and MAE, of six Kriging models trained by different physical subsets for each benchmark problem. Also, the reduction of test errors from physical Subsets 1 to 6 is calculated. Results showed that with the increase of physical training subset, the test errors are reduced greatly and the error reduction from Subsets 1 to 6 is near 90% and even 100%. These demonstrate the benefit of increasing the physical training dataset size on the surrogate model accuracy.

Table A. 1 Test errors of all surrogate models under six physical training subsets (1-6) of different problems

| Problem | Measure | Subset1 | Subset2 | Subset3 | Subset4 | Subset5 | Subset6 | Reduction (%) (Subset1→6) |
|---|---|---|---|---|---|---|---|---|
| *Wing Weight* | RMSE | 8.9289 | 7.2958 | 1.3745 | 0.9367 | 0.1715 | 0.1370 | -98.5 |
| | MAE | 6.8525 | 5.6023 | 1.0492 | 0.6649 | 0.1267 | 0.1021 | -98.5 |
| *U-beam* | RMSE | 0.0027 | 0.0012 | 0.0005 | 0.0002 | 0.0001 | 0.0000 | -99.3 |
| | MAE | 0.0020 | 0.0009 | 0.0003 | 0.0002 | 0.0000 | 0.0000 | -99.5 |
| *Pressure Vessel* | RMSE | 0.0446 | 0.0309 | 0.0196 | 0.0129 | 0.0097 | 0.0051 | -88.7 |
| | MAE | 0.0331 | 0.0227 | 0.0136 | 0.0085 | 0.0062 | 0.0032 | -90.2 |
| *Weld Beam* | RMSE | 6331.5730 | 69.2105 | 2.1566 | 0.1672 | 0.0479 | 0.0230 | -100.0 |
| | MAE | 4746.8262 | 49.8063 | 1.6514 | 0.1187 | 0.0292 | 0.0161 | -100.0 |

# Appendix B: The physical and pseudo dataset size effect

The effect of dataset sizes on the classification accuracy of four classification techniques for the $precision$, $TPR$, and $F1$ in Figure B. 1, Figure B. 2, and Figure B. 3, respectively. Similar trends are observed with the ACC as discussed in Section 5.3.

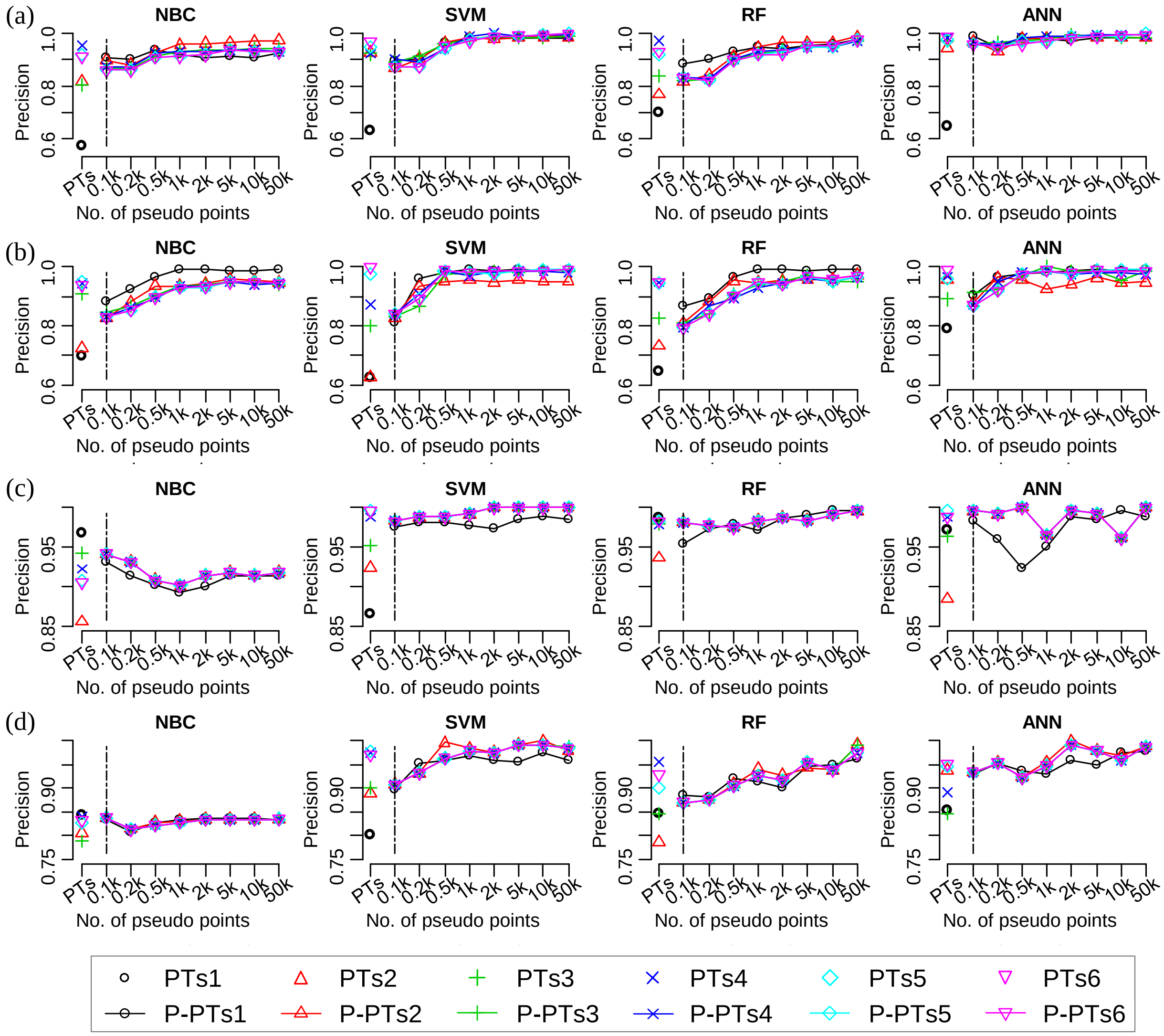

Figure B. 1 $precision$ of four ML models trained by different size pseudo datasets, which is compared with the $precision$ of the model trained by corresponding physical training subsets (PTs) for Problems (a) *Wing Weight*, (b) *U-beam*, (c) *Pressure Vessel*, and (d) *Weld Beam*

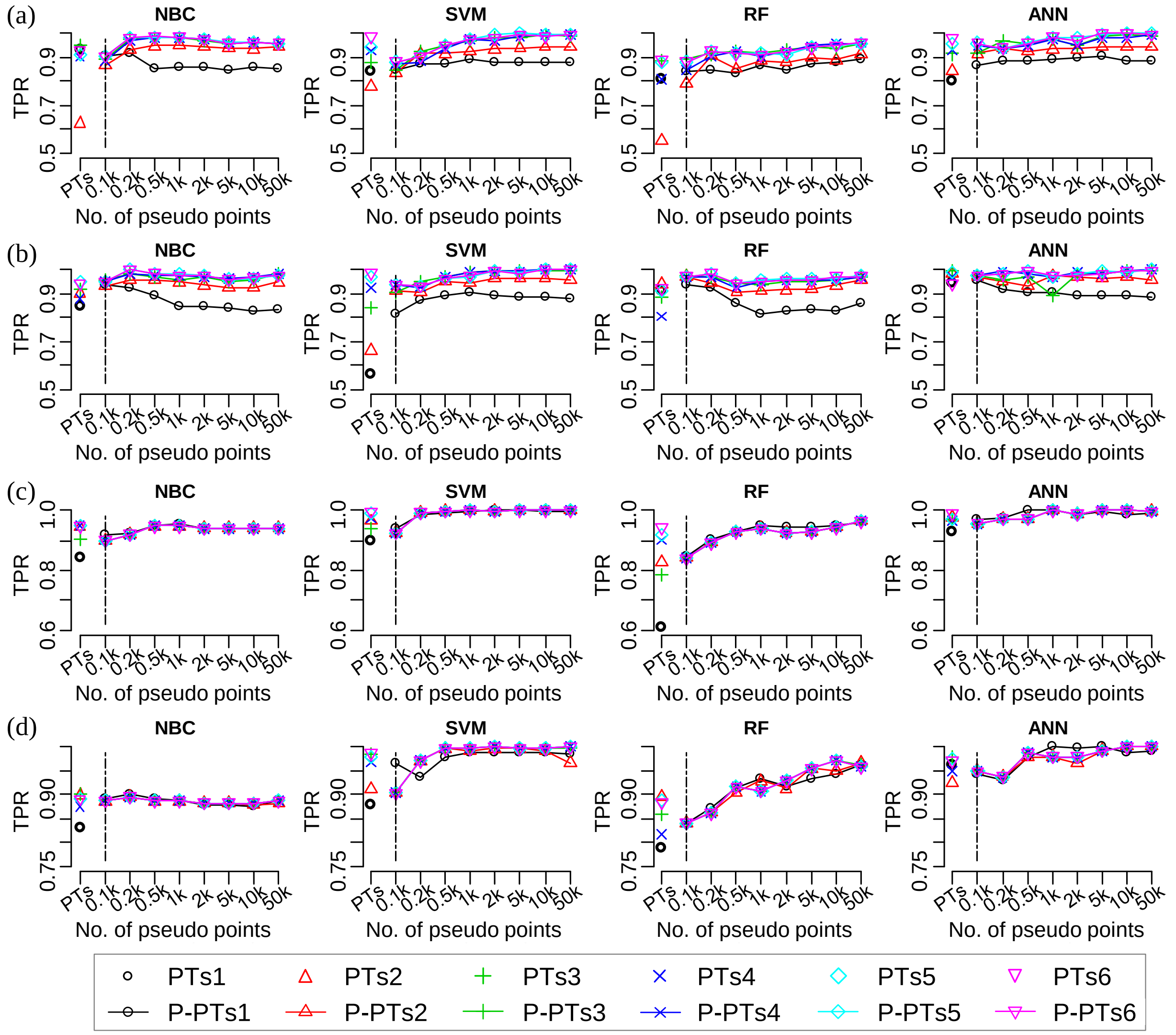

Figure B. 2 $TPR$ of four ML models trained by different size pseudo datasets, which is compared with the $TPR$ of the model trained by corresponding physical training subsets (PTs) for Problems (a) *Wing Weight*, (b) *U-beam*, (c) *Pressure Vessel*, and (d) *Weld Beam*

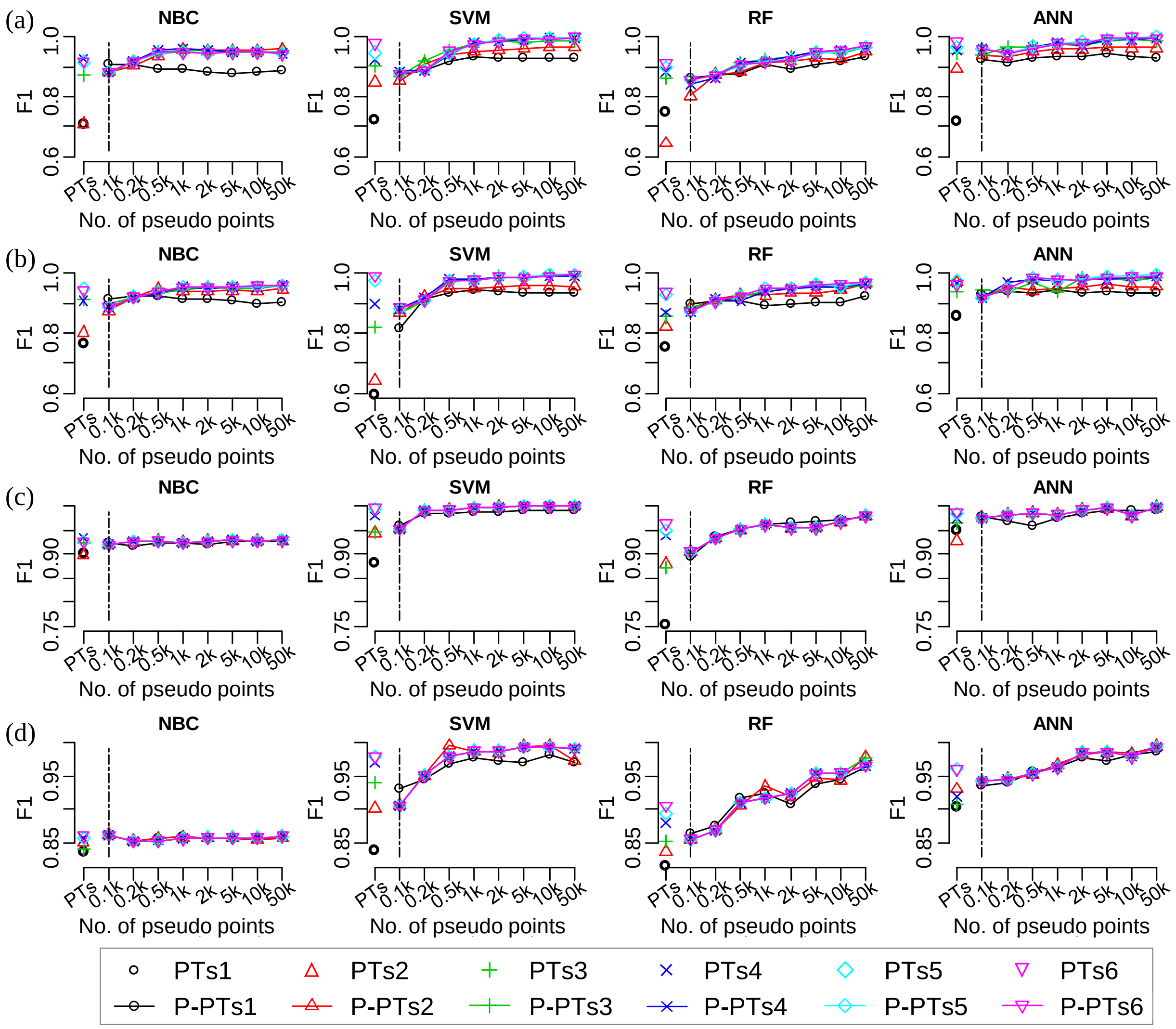

Figure B. 3 $F1$ of four ML models trained by different size pseudo datasets, which is compared with the $F1$ of the model trained by corresponding physical training subsets for Problems (a) *Wing Weight*, (b) *U-beam*, (c) *Pressure Vessel*, and (d) *Weld Beam*

# Appendix C: Surrogate and classification modeling for the DTU10MWPC problem

The physical training dataset of the DTU10MWPC trains a Kriging model. The resulted RMSE and MAE of the trained model and the fitting effect are shown in Figure C. 1(a), which suggests an accurate surrogate model. Then, 50,000 new combinations of $\lambda$ (TSR) and $\beta$ are generated and evaluated by the Kriging model, which forms the large pseudo dataset with the distribution of these two datasets shown in Figure C. 1(b). The physical datasets and pseudo datasets are then labeled: $C_p > 0.0$ for 'g', and otherwise, for 'p' for feasible domain recognition ('g'). This results in a 'g' and 'p' ratio of 1:1 approximately.

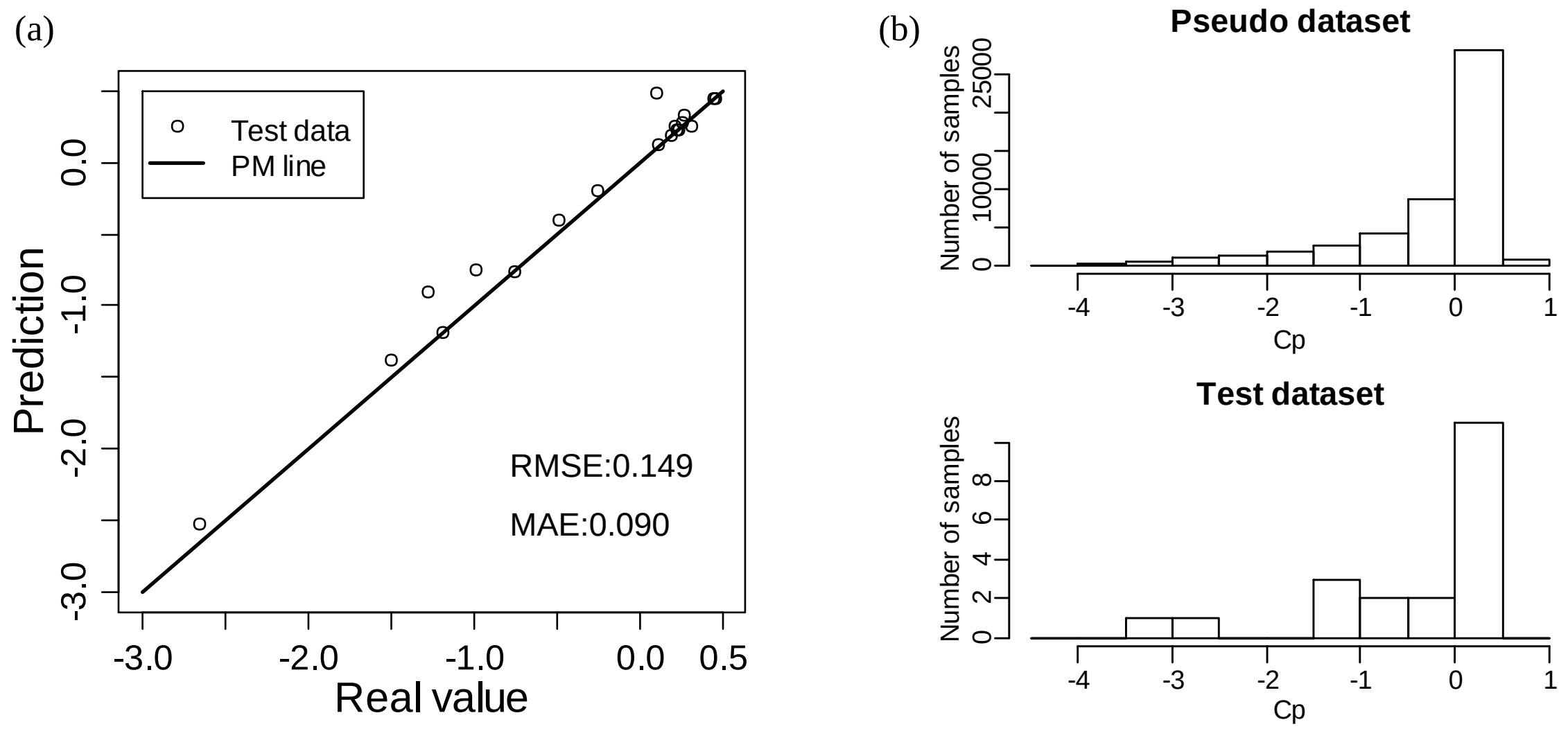

Figure C. 1 (a) Surrogate model prediction against real values using the physical test dataset with a PM line and (b) $C_p$ distribution of the pseudo and physical test datasets.

Hyperparameters of three methods except for NBC are also optimized. Using the labeled pseudo dataset, the NBC, SVM, RF, and ANN models are trained and compared with these trained by the labeled physical training dataset (20 points). The test accuracy is shown in Table C. 1. A great improvement can be observed by the large pseudo dataset.

Table C. 1 Accuracy of the four ML models trained by the small physical dataset and large pseudo dataset

| | NBC | | SVM | | RF | | ANN | |
|---|---|---|---|---|---|---|---|---|
| Measure | physical | pseudo | physical | pseudo | physical | pseudo | physical | pseudo |
| $ACC$ | 0.55 | 0.70 | 0.80 | 1.00 | 0.70 | 1.00 | 0.85 | 1.00 |
| $Precision$ | 0.58 | 0.73 | 0.77 | 1.00 | 0.86 | 1.00 | 0.79 | 1.00 |
| $TPR$ | 0.64 | 0.73 | 0.91 | 1.00 | 0.55 | 1.00 | 1.00 | 1.00 |
| $F1$ | 0.61 | 0.73 | 0.83 | 1.00 | 0.67 | 1.00 | 0.88 | 1.00 |